\documentclass[submission, PhysProc]{SciPost}

\hypersetup{
    colorlinks,
    linkcolor={red!50!black},
    citecolor={blue!50!black},
    urlcolor={blue!80!black}
}

\usepackage[bitstream-charter]{mathdesign}
\usepackage{cite}

\DeclareSymbolFont{usualmathcal}{OMS}{cmsy}{m}{n}
\DeclareSymbolFontAlphabet{\mathcal}{usualmathcal}

\begin{document}

\begin{center}{\Large \textbf{
Description of continuum structures in a discrete basis: Three-body resonances and two-nucleon decays\\
}}\end{center}

\begin{center}
J. Casal\textsuperscript{1,2$\star$},
M. Rodríguez-Gallardo\textsuperscript{3,4},
J. M. Arias\textsuperscript{3,4},
J. Gómez-Camacho\textsuperscript{3,5},\\
L. Fortunato\textsuperscript{1,2} and
A. Vitturi\textsuperscript{1,2}
\end{center}

\begin{center}
{\bf 1} Dipartimento di Fisica e Astronomia ``G.~Galilei'', Università degli Studi di Padova, via Marzolo 8, I-35131 Padova, Italy
\\
{\bf 2} INFN - Sezione di Padova, via Marzolo 8, I-35131 Padova, Italy
\\
{\bf 3} Departamento de Física Atómica, Molecular y Nuclear, Facultad de Física, Universidad de Sevilla, Apdo. 1065, E-41080 Sevilla, Spain
\\
{\bf 4} Instituto Carlos I de Física Teórica y Computacional, Universidad de Sevilla, Sevilla, Spain\\
{\bf 5} Centro Nacional de Aceleradores, U. Sevilla, J. Andalucía, CSIC, Tomas A. Edison 7, E-41092 Sevilla, Spain
\\
${}^\star$ {\small \sf casal@pd.infn.it}
\end{center}

\begin{center}
\today
\end{center}


\definecolor{palegray}{gray}{0.95}
\begin{center}
\colorbox{palegray}{
  \begin{tabular}{rr}
  \begin{minipage}{0.05\textwidth}
    \includegraphics[width=14mm]{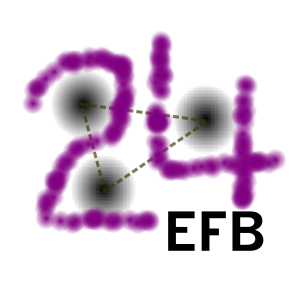}
  \end{minipage}
  &
  \begin{minipage}{0.82\textwidth}
    \begin{center}
    {\it Proceedings for the 24th edition of European Few Body Conference,}\\
    {\it Surrey, UK, 2-6 September 2019} \\
    \end{center}
  \end{minipage}
\end{tabular}
}
\end{center}
\vspace{-25pt}

\section*{Abstract}
{\bf
Weakly bound and unbound three-body nuclei are studied by using the pseudostate method within the hyperspherical formalism. After introducing the theoretical framework, the method is applied first to the $\boldsymbol{^9}$Be nucleus, showing a good agreement with the available data for its low-lying dipole response. Then, recent results on the structure and decay of the two-neutron emitters $\boldsymbol{^{26}}$O and $\boldsymbol{^{16}}$Be are presented. In particular, the role of the $\boldsymbol{n}$-$\boldsymbol{n}$ correlation in shaping their properties is discussed. 
}

\noindent\rule{\textwidth}{1pt}
\tableofcontents\thispagestyle{fancy}
\noindent\rule{\textwidth}{1pt}

\section{Introduction}
\label{sec:intro}
Recent advances in radioactive ion beam physics and detection techniques have triggered the exploration of the exotic properties and decay modes of light nuclear systems at the limit of stability and beyond the driplines. Large experimental and theoretical efforts have been devoted to understanding the structure and reaction dynamics of loosely bound systems, such as halo nuclei, where continuum effects are of utmost relevance~\cite{tanihata13}. Of particular interest is the case of two-neutron halo nuclei, e.g., $^6$He, $^{11}$Li or $^{14}$Be. These $\text{core}+N+N$ nuclei are Borromean, i.e., three-body systems in which all binary subsystems cannot form bound states. While the correlations between the valence neutrons are known to play a fundamental role in shaping the properties of two-neutron halo nuclei~\cite{kikuchi16}, a proper understanding of their structure requires also solid constrains on the unbound binary subsystems $^5$He, $^{10}$Li or $^{13}$Be~\cite{aksyutina13a}. The evolution of these correlations beyond the driplines gives rise to two-neutron emitters, e.g., $^{16}$Be or $^{26}$O~\cite{spyrou12,kohley13}. A similar situation can be found for proton-rich nuclei. For instance, the Borromean $^{17}$Ne nucleus, characterized by the properties of its unbound subsystem $^{16}$F, has been proposed to exhibit a two-proton halo, while other exotic systems, such as $^{6}$Be and $^{11}$O (the mirror nuclei of $^6$He and $^{11}$Li, respectively), are two-proton emitters~\cite{oishi17}. Since they have a marked $\text{core}+N+N$ character, three-body models are a natural choice to describe their structure and processes involving them~\cite{zhukov93}. The description of the continuum in three-body nuclei, however, is not an easy task. 

In this contribution, we present some recent developments in the description of weakly bound and unbound three-body nuclei. First, we introduce the hyperspherical framework, and we briefly discuss the pseudostate method and its possible applications. As an example, we study the dipole response of $^9$Be in a three-body $\alpha+\alpha+n$ model. Then, we recall different approaches to identify and characterize few-body resonances in a discrete basis, and we focus on the case of the two-neutron unbound systems $^{26}$O and $^{16}$Be using a $\text{core}+n+n$ representation. Finally, the main conclusions of this work are summarized.

\section{Hyperspherical formalism for three-body systems}
\begin{figure}[t]
	\centering
	\includegraphics[width=0.225\linewidth]{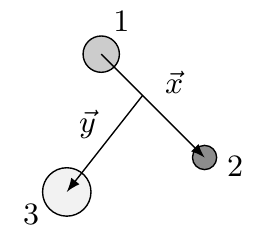}
	\caption{Jacobi coordinates for a three-body system.}
	\label{fig:jacobi}
\end{figure}

Three-body systems, such as the Borromean stable nuclei $^{9}$Be ($\alpha+\alpha+n$) and $^{12}$C ($\alpha+\alpha+\alpha$), the exotic two-neutron halo $^{6}$He ($\alpha+n+n$) and $^{11}$Li ($^9\text{Li}+n+n$), or the unbound two-nucleon emitters $^{6}$Be ($\alpha+p+p$) and $^{26}$O ($^{24}\text{O}+n+n$), can be described using the hyperspherical harmonics (HH) framework~\cite{zhukov93,nielsen01}. In this approach, Hamiltonian eigenstates are expanded as
\begin{equation}
\Psi^{j\mu}(\rho,\Omega) = \rho^{-5/2}\sum_{\beta}R_{\beta }^j(\rho)\mathcal{Y}_{\beta}^{j\mu}(\Omega),
\label{eq:3bwf}
\end{equation}
where the hyperspherical coordinates $\displaystyle\rho=\sqrt{x^2+y^2}$ and $\displaystyle\Omega=\{\alpha,\widehat{x},\widehat{y}\}$, $\displaystyle\tan\alpha=x/y$, are introduced. Here, the hyper-radius ($\rho$) and the hyperangle ($\alpha$) are defined from the scaled Jacobi coordinates $\{\boldsymbol{x},\boldsymbol{y}\}$ depicted in Fig.~\ref{fig:jacobi}. These are related to physical distances by
\begin{equation}
\boldsymbol{x}=\boldsymbol{r}_x\sqrt{\frac{A_1A_2}{A_1+A_2}}, ~~~~~~~\boldsymbol{y}=\boldsymbol{r}_y\sqrt{\frac{A_3(A_1+A_2)}{A_1+A_2+A_3}}.
\label{eq:jacphys}
\end{equation}
The index $\beta\equiv\{K,l_x,l_y,l,S_x,j_{ab}\}$ in Eq.~(\ref{eq:3bwf}) represents the channels coupled to total angular momentum $j$, so that $R_{\beta}^{j}(\rho)$ is the radial part for each one, and $\mathcal{Y}_{\beta}^{j\mu}(\Omega)$ follows the coupling order
\begin{equation}
\mathcal{Y}_{\beta}^{j\mu}(\Omega)=\left\{\left[\Upsilon_{Klm_l}^{l_xl_y}(\Omega)\otimes\kappa_{s_x}\right]_{j_{ab}}\otimes\phi_I\right\}_{j\mu}.
\label{eq:Yfun}
\end{equation}
In the previous expression, $l_x$ and $l_y$ are the orbital angular momenta associated to $\boldsymbol{x}$ and $\boldsymbol{y}$, respectively, and $\boldsymbol{l}=\boldsymbol{l}_x+\boldsymbol{l}_y$; $S_x$ is the total spin of the particles related by $\boldsymbol{x}$, so that $\boldsymbol{j}_{ab}=\boldsymbol{l}+\boldsymbol{S}_x$; and the total angular momentum $j$ is obtained by coupling $j_{ab}$ with the spin of the third particle $I$ (which is assumed to be fixed), i.e., $\boldsymbol{j}=\boldsymbol{j}_{ab}+\boldsymbol{I}$. Equation~(\ref{eq:Yfun}) is written in terms of hyperspherical harmonics (HH) $\Upsilon_{Klm_l}^{l_xl_y}$, which are the eigenfunctions of the hypermomentum operator $\widehat{K}$ and follow the analytical form
\begin{equation}
\Upsilon_{Klm_l}^{l_xl_y}(\Omega)=\varphi_K^{l_xl_y}(\alpha)\left[Y_{l_x}(\boldsymbol{x})\otimes Y_{l_y}(\boldsymbol{y})\right]_{lm_l},
\label{eq:HH}
\end{equation}
\begin{equation}
\varphi_K^{l_xl_y}(\alpha) = N_{K}^{l_xl_y}\left(\sin\alpha\right)^{l_x}\left(\cos\alpha\right)^{l_y} P_n^{l_x+\frac{1}{2},l_y+\frac{1}{2}}\left(\cos 2\alpha\right),
\label{eq:varphi}
\end{equation}
with $P_n^{a,b}$ a Jacobi polynomial of order $n=(K-l_x-l_y)/2$ and $N_K^{l_xl_y}$ a normalization constant. Using the above definitions, and inserting the expansion~(\ref{eq:3bwf}) in the Schr\"odinger equation, the problem reduces to solving a set of coupled equations in the hyperradial coordinate,
\begin{equation}
\left[-\frac{\hbar^2}{2m}\left(\frac{d^2}{d\rho^2}-\frac{15/4+K(K+4)}{\rho^2}\right)-\varepsilon\right]R_{\beta}^{j}(\rho)+\sum_{\beta'}V_{\beta'\beta}^{j\mu}(\rho) R_{\beta'}^{j}(\rho)=0,
\label{eq:coupled}
\end{equation}
involving the coupling potentials
\begin{equation}
V_{\beta'\beta}^{j\mu}(\rho)=\left\langle \mathcal{Y}_{\beta }^{j\mu}(\Omega)\Big|V_{12}+V_{13}+V_{23} \Big|\mathcal{Y}_{\beta'}^{ j\mu}(\Omega) \right\rangle.
\label{eq:coup}
\end{equation}

\section{The Pseudostate method}
Equation~(\ref{eq:coupled}) can be solved for negative (bound) or positive (scattering) energy eigenstates by imposing standard boundary conditions. The latter, however, is computationally challenging because the expansion~(\ref{eq:3bwf}) for scattering states requires a combination of incoming and outgoing channels (see, e.g., Ref.~\cite{lovell17}). Moreover, for systems comprising several charged particles the matching with the asymptotics poses additional challenges due to the long-range nature of the Coulomb interaction~\cite{nguyen12}. An alternative to this approach is the so-called pseudo state (PS) method~\cite{tolstikhin97}, in which the true continuum is approximated by a set of discretized, normalizable states. In this method, the set of coupled equations is replaced by a standard eigenvalue problem, so the hyperradial functions are expanded in a given basis,
\begin{equation}
R_{\beta}^{nj}(\rho)=\sum_i D_{i\beta}^{nj}U_{i\beta}(\rho).
\label{eq:PS}
\end{equation}
Here, index $i$ counts the hyperradial excitations up to $N$ (i.e., the number of basis functions included for each channel), and $D_{i\beta}^{nj}$ coefficients are obtained by diagonalizing the three-body Hamiltonian in the basis $\{U_{i\beta}\}$. Note the new index $n$ which labels the finite number of states, $\Psi_n^{j\mu}$, associated to discrete energies $\varepsilon_n$. Eigenstates corresponding to negative-energy eigenvalues describe the bound states of the system, and those with positive energies provide a discrete representation of the continuum.

\begin{figure}[t]
	\centering
	\includegraphics[width=0.475\linewidth]{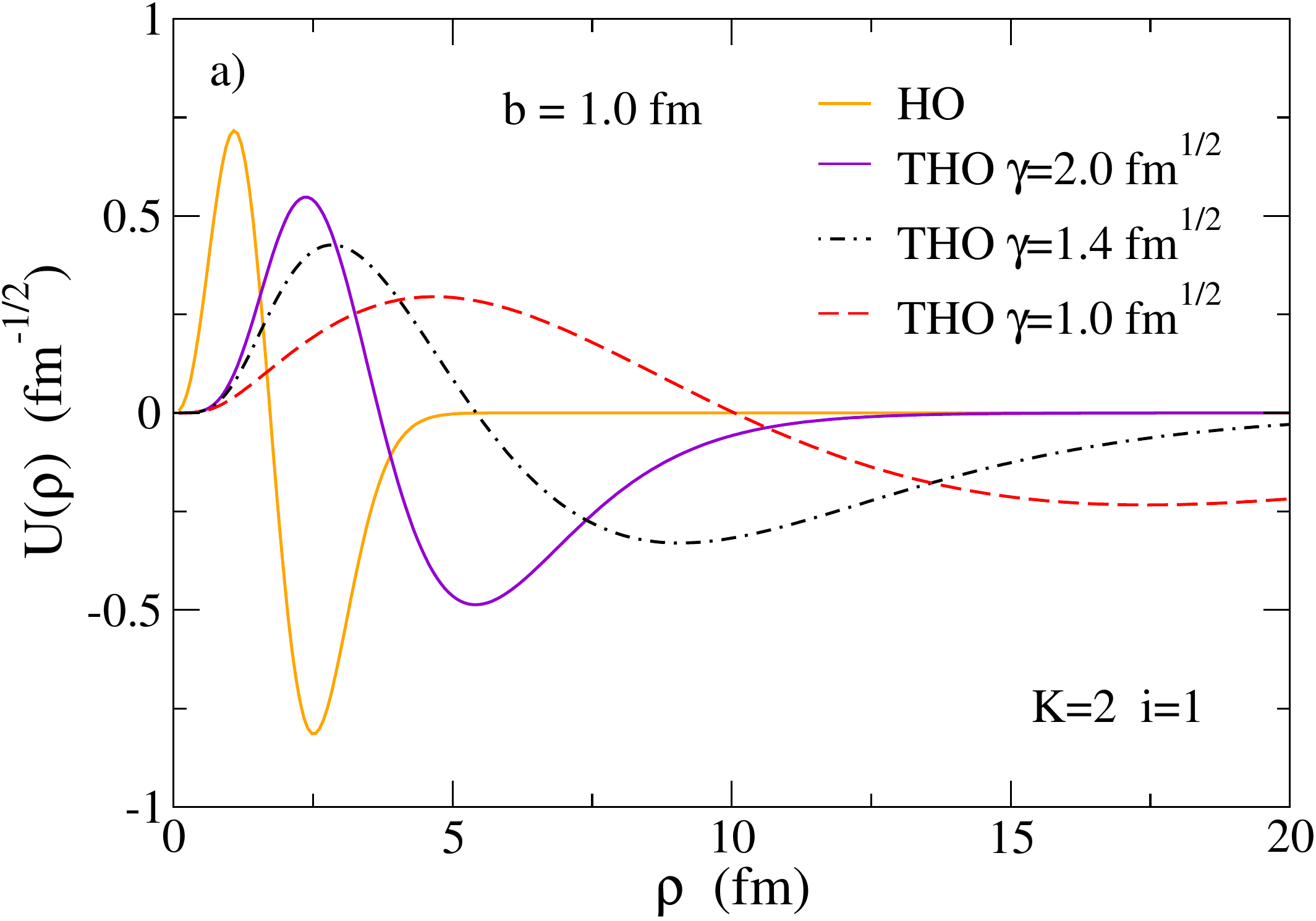}\hspace{10pt}\includegraphics[width=0.475\linewidth]{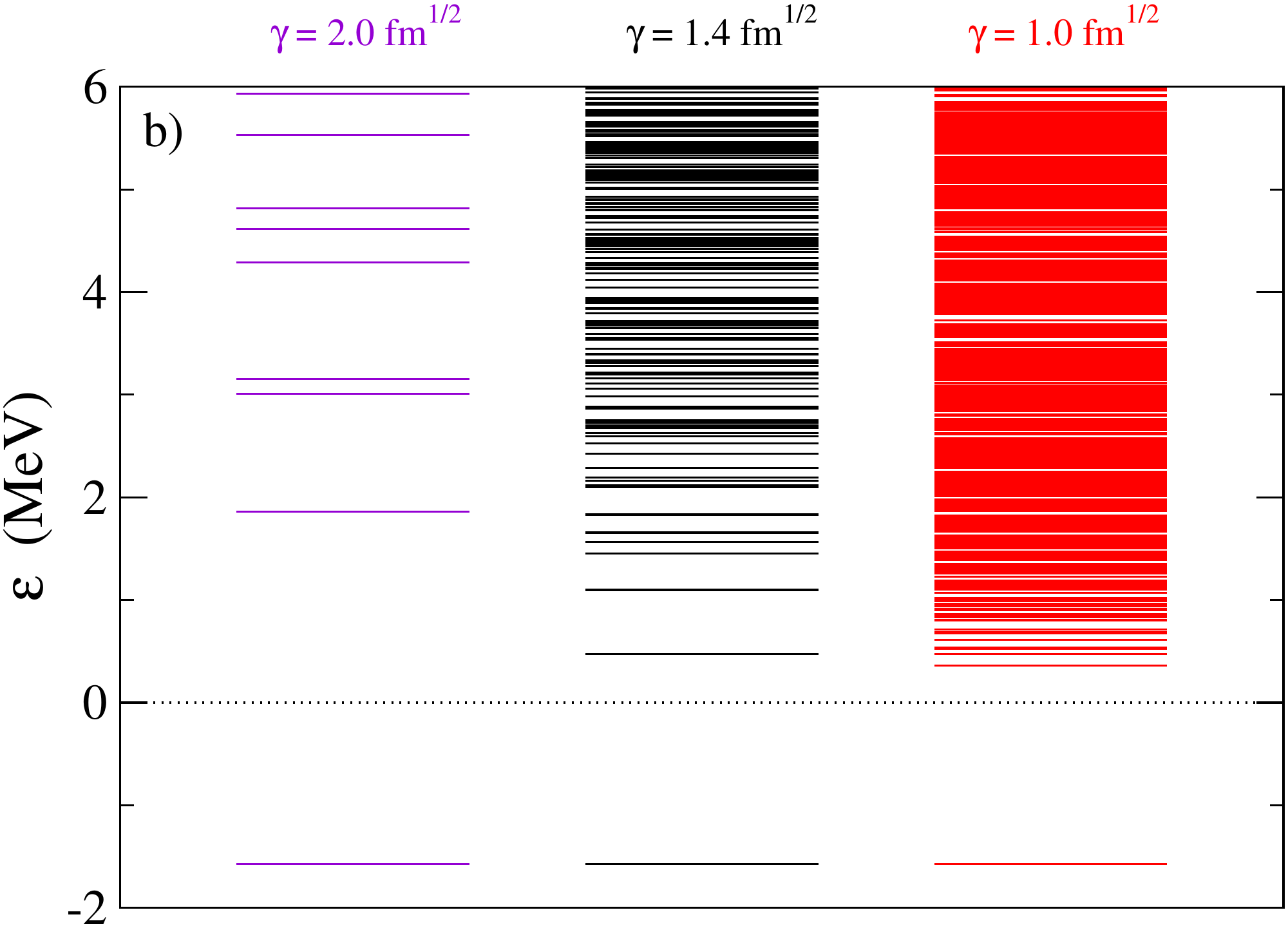}
	\caption{a) THO basis functions, obtained with $b=1$ fm and different values of the $\gamma$ parameter, compared with a standard HO function for a given channel component ($K=2,~i=1$). b) Resulting PS spectra up to 6 MeV with respect to the three-body threshold, for a system characterized by a single bound state, obtained with three different THO basis with the same values of $\gamma$ shown in the left panel.}
	\label{fig:hotho}
\end{figure}

Different choices for the basis functions have been explored in the literature~\cite{desc03,matsumoto04,mroga05}. In this work, we use the analytical transformed harmonic oscillator (THO) basis~\cite{casal14}. The basis functions are built from the harmonic oscillator (HO) ones by performing a local scale transformation,
\begin{equation}
U_{i\beta}^{\text{THO}}(\rho)=\sqrt{\frac{ds}{d\rho}}U_{iK}^{\text{HO}}[s(\rho)],
\label{eq:HOTHO}
\end{equation}
with the condition that the asymptotic behavior show an exponential decay, rather than a $e^{-\rho^2}$ behavior. This can be achieved by using the analytical transformation proposed in Ref.~\cite{karataglidis05}, 
\begin{equation}
s(\rho) = \frac{1}{\sqrt{2}b}\left[\frac{1}{\left(\frac{1}{\rho}\right)^{4} +
	\left(\frac{1}{\gamma\sqrt{\rho}}\right)^4}\right]^{\frac{1}{4}},
\label{eq:LST}
\end{equation}
depending on parameters $b$ and $\gamma$. An interesting feature of this transformation is that the ratio $\gamma/b$ governs the asymptotic behavior of the basis functions, thus changing the density of PS, after diagonalization, as a function of the energy. This is illustrated in Fig.~\ref{fig:hotho}, where we show that a smaller $\gamma$ parameter (for a fixed oscillator length $b$) provides basis functions with larger hyperradial extension and tends to concentrate more PS in the energy region close to the breakup threshold. In some cases, one is interested in this kind of dense, detailed representation of the continuum, for instance to compute decay-energy spectra or electromagnetic transition probability distributions. In other cases, however, one can choose a THO basis with a smaller hyperradial extension (i.e., larger $\gamma$), which leads to a continuum representation characterized by a small number of states close to the breakup threshold. As shown in Ref.~\cite{casal18}, this can be used to identify three-body resonances following the so-called stabilization method~\cite{HaziTaylor70,TaylorHazi76}. Examples on these two approaches will be shown in the following sections.

\subsection{Example: Dipole response in $^9$Be using PS}
As an example of the three-body description of weakly bound systems using PS within the hyperspherical framework, we consider the case of the Borromean nucleus $^{9}$Be ($\alpha+\alpha+n$), which has previously attracted remarkable theoretical and experimental attention. On the one hand, $^9$Be is stable, but has a small binding energy of the last neutron, so continuum effects have been shown to be important in describing low-energy nuclear reactions~\cite{Descouvemont15} involving this nucleus. On the other hand, its formation via the $\alpha(\alpha n,\gamma){^9}$Be reaction followed by $^{9}\text{Be}(\alpha,n){^{12}}$C has been linked to the $r$-process nucleosynthesis~\cite{Sumiyoshi02,casal14}. In this context, the structure properties of its ground state and low-lying resonances play a key role. 

In Ref.~\cite{casal14}, we presented a study of $^9$Be by using the PS method in hyperspherical coordinates, focusing on the properties of its ground state and its low-lying E1 and M1 continuum. Three-body $\alpha+\alpha+n$ states for total angular momenta $j^\pi=1/2^\pm,3/2^\pm,5/2^\pm$ were described in a THO basis, with effective $n$-$n$ and $\alpha$-$n$ interactions as main ingredients. 

For the 3/2$^-$ ground state, converged calculations were achieved by including hypermomenta up to $K_{max}=30$ and $N=20$ hyperradial excitations in each channel. A THO basis with $b=0.7$ fm and $\gamma=1.4$ fm$^{1/2}$ was employed, although results for the ground state were independent on this choice. An additional small three-body force was needed in Eq.~(\ref{eq:coup}) to recover the ground-state energy of -1.574 MeV. The computed charge radii ($r_{ch}=2.508$ fm) and quadrupole moment ($Q_2=4.91$ efm$^2$) were in rather good agrement with the available data. The ground-state wave function was found to contain $^8\text{Be}(g.s.)+n$ and $^{8}\text{Be}(2^+)+n$ configurations with 52.5\% and 46.6\% of the total norm, respectively, pointing towards large core excitations in effective two-body models for $^9$Be.

\begin{figure}[t]
	\centering
	\includegraphics[width=0.475\linewidth]{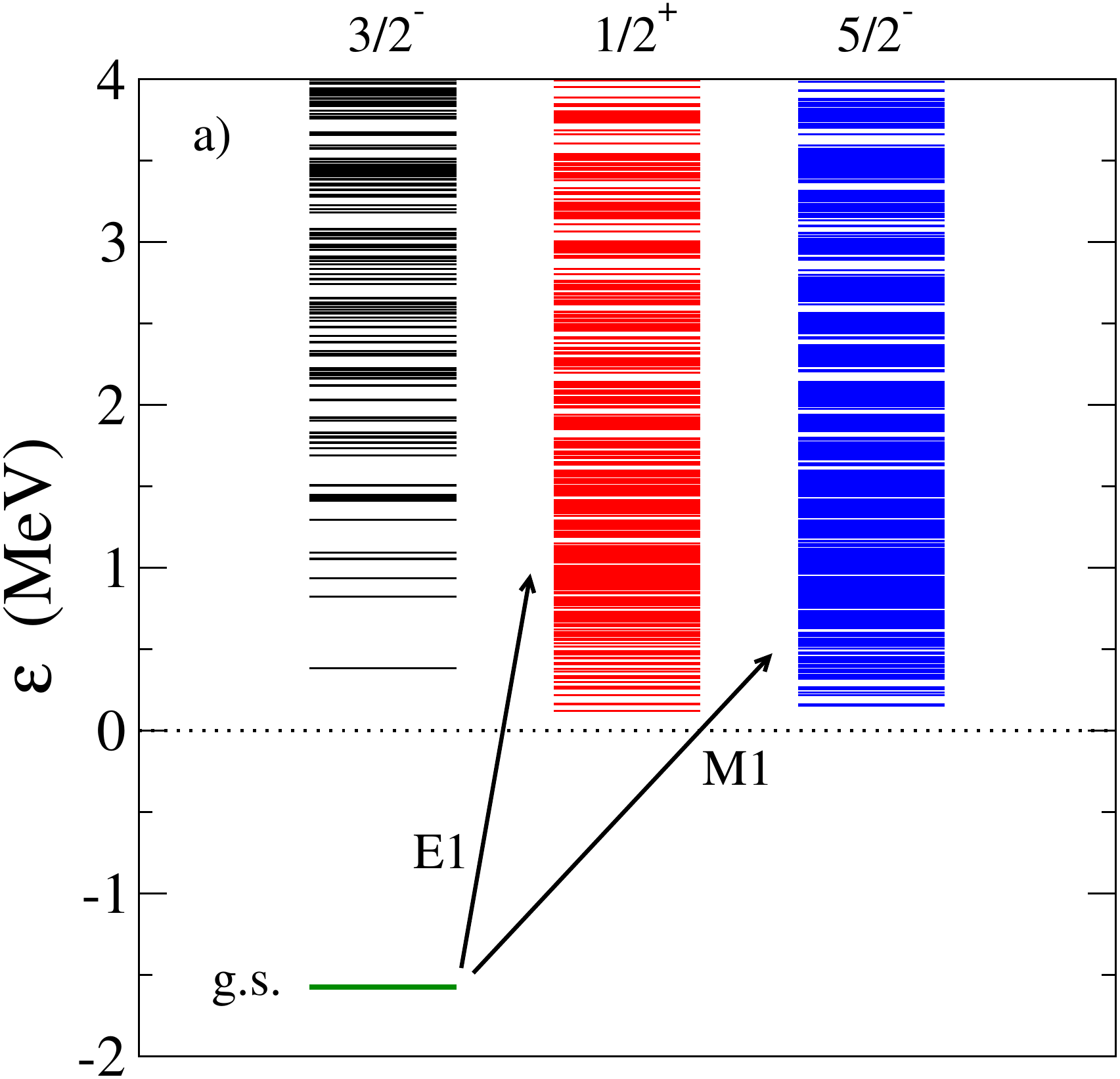}\hspace{15pt}\includegraphics[width=0.475\linewidth]{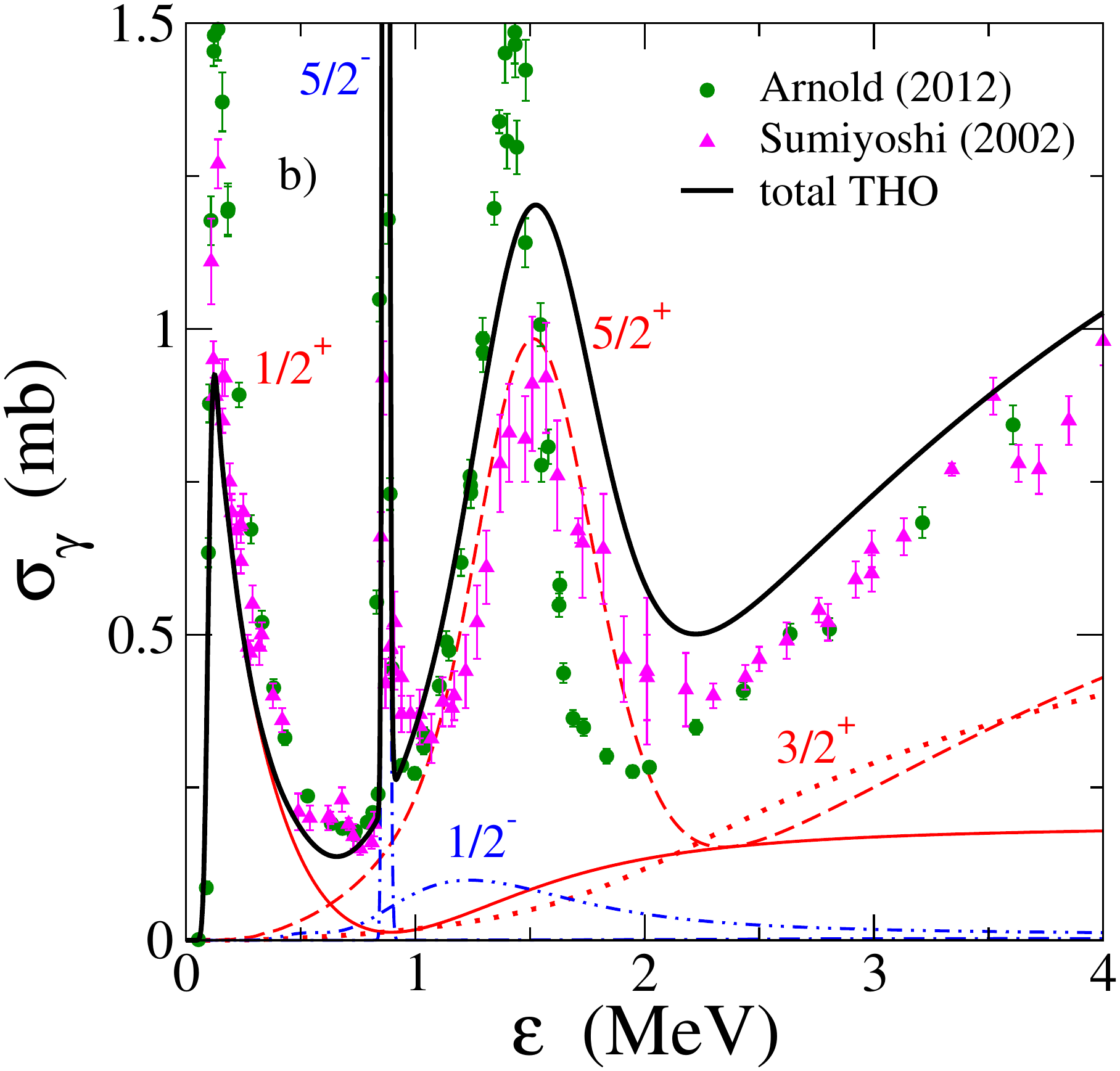}
	\caption{a) Spectra for 3/2$^-$, 1/2$^+$ and 5/2$^-$ states in $^9$Be up to 4 MeV above the $\alpha+\alpha+n$ threshold. b) Photodissociation cross section for $^9$Be resulting from E1 (red) and M1 (blue) transitions, including also the contribution to 3/2$^+$, 5/2$^+$ and 1/2$^-$ states, compared with the data from Refs.~\cite{Sumiyoshi02,Arnold12}.}
	\label{fig:E1M1}
\end{figure}

Details on the calculations for 1/2$^+$, 3/2$^+$, 5/2$^+$, 1/2$^-$ and 5/2$^-$ states can be found in Ref.~\cite{casal14}. In these cases, THO bases with smaller $\gamma$ values were used to increase the PS density close to the breakup threshold, ensuring a good representation of the low-energy continuum. From the matrix elements of the E1 and M1 operators, we computed the corresponding electromagnetic dipole transitions from the 3/2$^-$ ground state and states with different angular momenta (illustrated in Fig.~\ref{fig:E1M1}a), looking in particular for the appearance of resonant peaks. In Fig.~\ref{fig:E1M1}b, we present the computed photodissociation cross section for $^9$Be, obtained from the electromagnetic transition probability distributions ($\mathcal{O}=E$ or $M$) through~\cite{RdDiego10,Forseen03}
\begin{equation}
\sigma_\gamma^{(\mathcal{O}\lambda)}(\varepsilon_\gamma)=\frac{(2\pi)^3 (\lambda+1)}
{\lambda[(2\lambda+1)!!]^2}\left(\frac{\varepsilon_\gamma}{\hbar
	c}\right)^{2\lambda-1}\frac{dB(\mathcal{O}\lambda)}{d\varepsilon},
\end{equation}
and compared with two different sets of experimental data~\cite{Sumiyoshi02,Arnold12}. Our calculations are able to describe the data rather well, in particular reproducing the low-lying peaks arising from the 1/2$^+$ and 5/2$^-$ resonances in $^9$Be. In addition, four-body continuum-discretized coupled-channels (CDCC) calculations using our set of PS were also found to provide a good representation of the continuum for the scattering of $^9$Be on heavy ions at near-barrier energies~\cite{casal15}. Note that our approach treats resonant and non-resonant states on the same footing, thus confirming the suitability of the PS approach within the hyperspherical framework to describe bound and continuum states in weakly bound three-body nuclei. 

\subsection{Three-body decay energy in $^{26}$O}
The $^{26}$O nucleus was observed for the first time in a proton-knockout reaction from $^{27}$F, its ground state being located only a few keV above the $^{24}\text{O}+n+n$ threshold~\cite{Lunderberg12,kohley13}. A more recent experiment reported this 0$^+$ ground-state resonance energy to be 18 keV, together with a second state (likely the first 2$^+$) at 1.28 MeV~\cite{Kondo16}. Since $^{25}$O is also unbound, characterized by a low-lying $d_{3/2}$ resonance at 750 keV~\cite{Kondo16}, the sequential decay of the $^{26}$O nucleus is energetically unavailable, bringing interest in the context of two-nucleon decays and two-neutron radioactivity. A theoretical investigation of the decay dynamics of $^{26}$O within a $^{24}\text{O}+n+n$ model, using the Green's function method to study the three-body decay energy spectrum, was recently shown to describe rather well the available data~\cite{Hagino16}. In that work, a $^{24}\text{O}+n$ potential was adjusted to reproduce the energy of the $d_{3/2}$ ground-state resonance in $^{25}$O. In addition, they employed a density-dependent pairing interaction, whose depth was fixed to reproduce the two-neutron separation energy in $^{26}$O. 

\begin{figure}[t]
	\centering
	\includegraphics[width=0.475\linewidth]{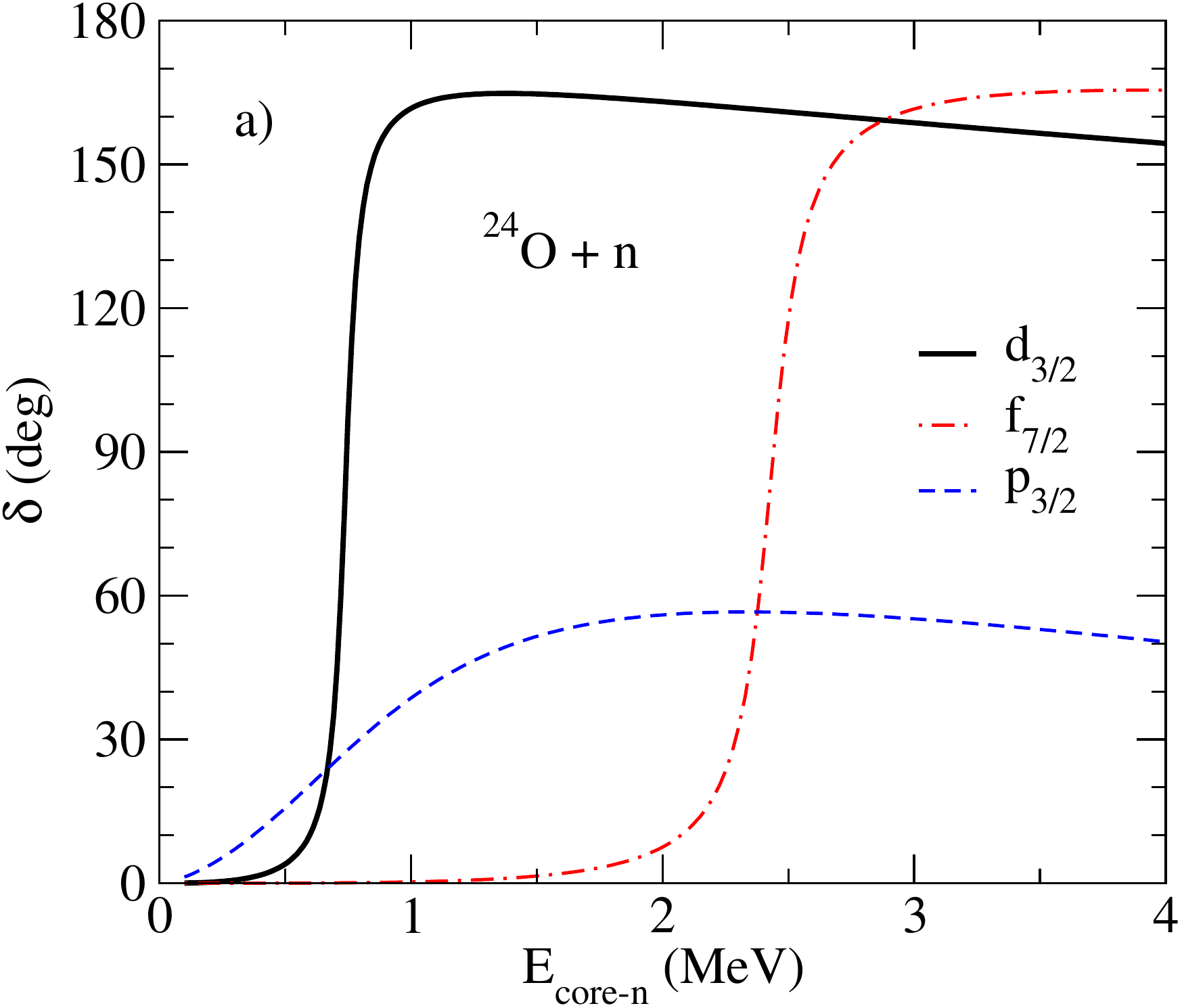}\hspace{15pt}\includegraphics[width=0.475\linewidth]{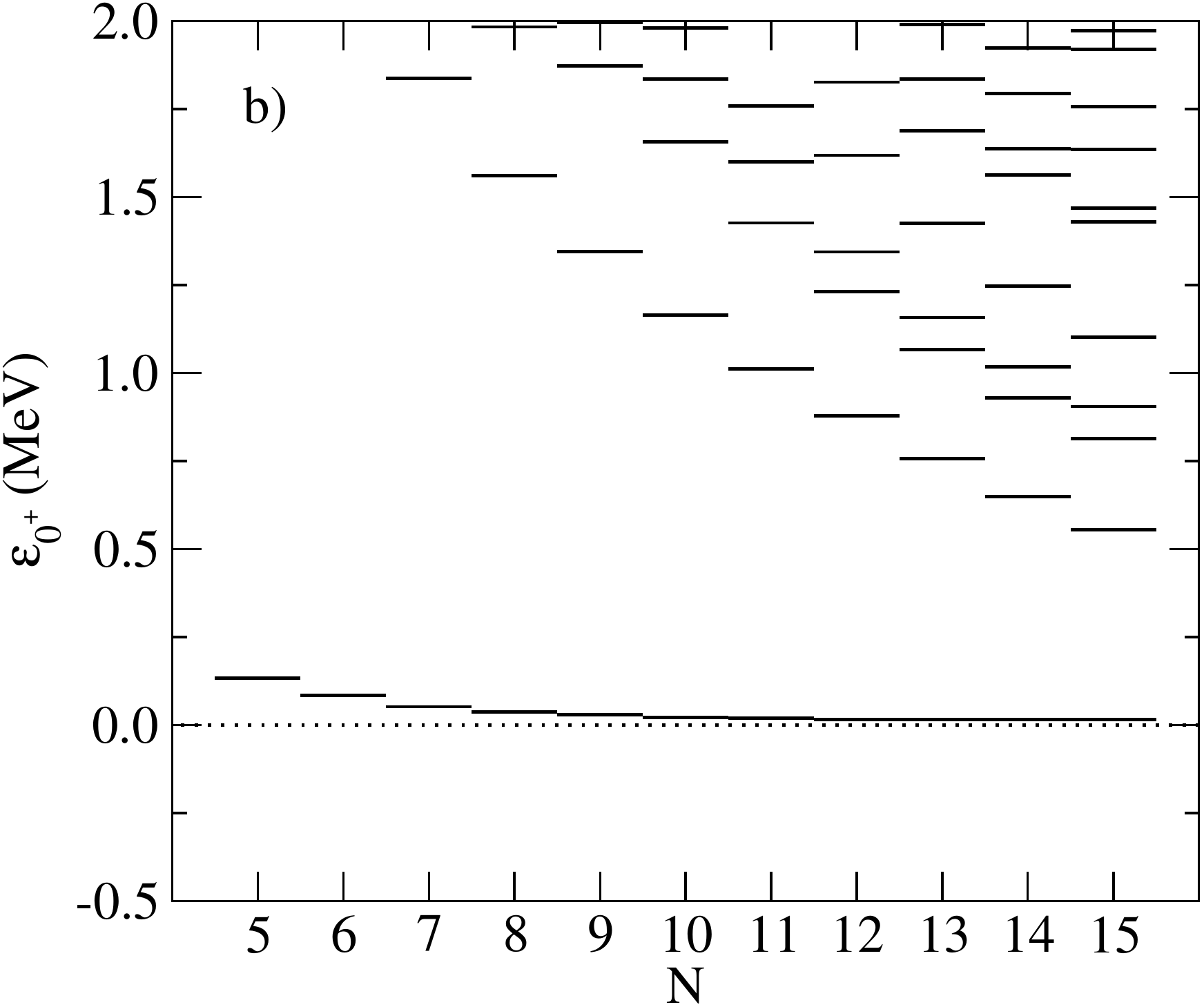}
	\caption{a) $^{24}\text{O}+n$ phase shifts using the potential from Ref.~\cite{Hagino16}. b) PS spectra for 0$^+$ states in $^{26}$O as a function of the number of basis functions.}
	\label{fig:26O}
\end{figure}

\begin{figure}[t]
	\centering
	\includegraphics[width=0.225\linewidth]{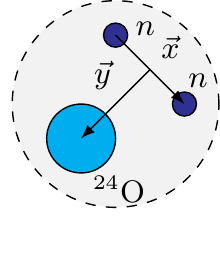}\hspace{-25pt}\includegraphics[width=0.65\linewidth]{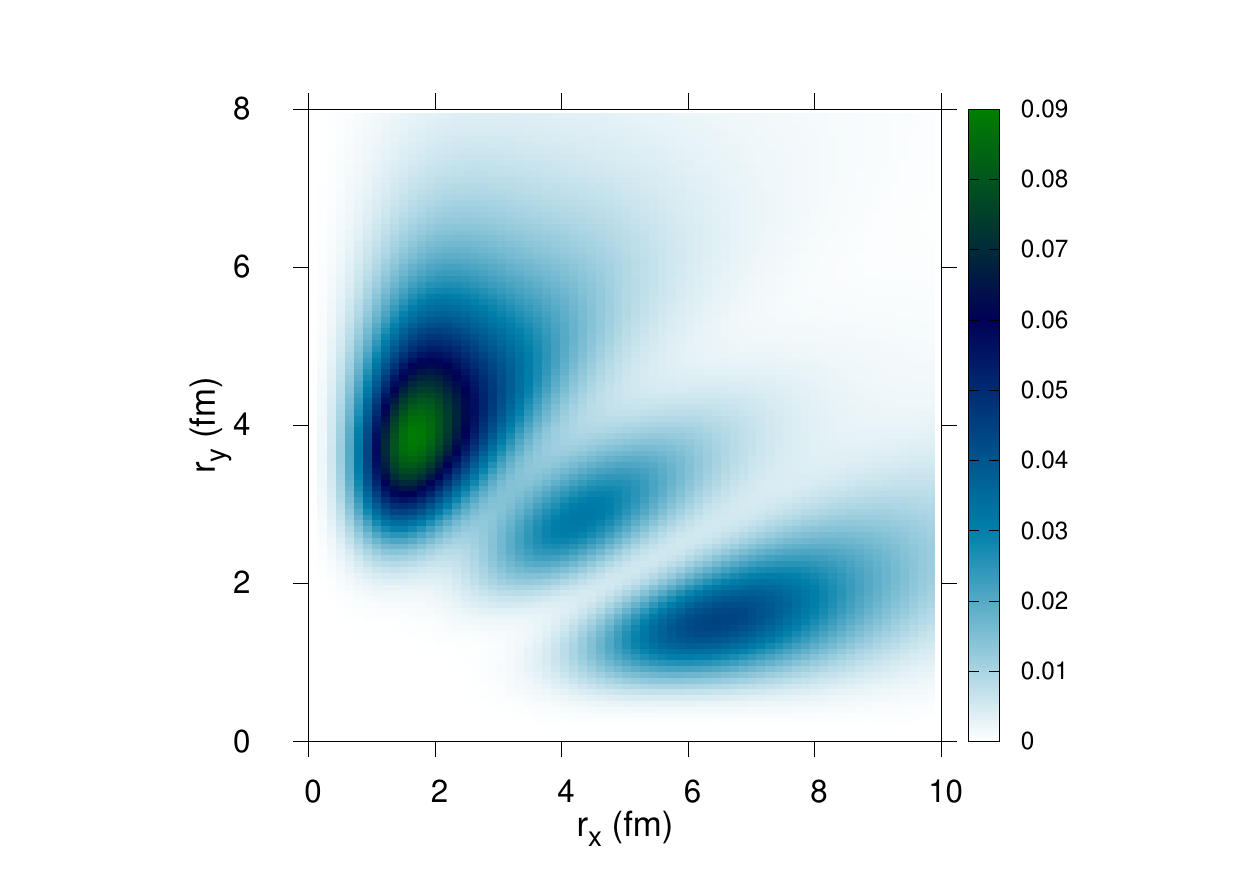}
	\caption{Probability density (fm$^{-2}$) for the 0$^+$ ground-state resonance in $^{26}$O (right panel) as a function of the distances defined by the Jacobi coordinates (left panel)}
	\label{fig:26Onn}
\end{figure}

In the present work, we study the properties of $^{26}$O and its decay energy spectrum after one-proton removal from $^{27}$F by using the PS method. For that purpose, we use the same $^{24}\text{O}+n$ potential developed in Ref.~\cite{Hagino16}. The corresponding phase shifts for $d_{3/2}$, $f_{7/2}$ and $p_{3/2}$ states are shown in Fig.~\ref{fig:26O}a. For the interaction between the valence neutrons, we employ the GPT $n$-$n$ parametrization~\cite{GPT}. Instead of fixing the depth of this interaction, we add a phenomenological three-body force which acts as a small correction on the spectrum to reproduce the 0$^+$ energy. 
However, as discussed above, the PS method provides different representations of the continuum depending on the basis choice. Therefore, a question arises on how to identify the ground-state resonance within this approach. In this case, since the $0^+$ state in $^{26}$O is almost bound and is characterized by a extremely small width, its properties can be described reasonably by using the so-called stabilization approach~\cite{HaziTaylor70,TaylorHazi76}. This implies that stable, square-integrable eigenstates close to the resonance energy provide a good representation of the inner part of the exact scattering wave function. The stability of the states within the PS approach can be analyzed, for instance, by studying the evolution of the spectrum with the number of basis functions $N$. Preliminary results are shown in Fig.~\ref{fig:26O}b, where the lowest eigenstate converges rather fast to the resonance energy. These calculations correspond to $K_{max}=20$ in the hyperspherical expansion. The computed spectra have been obtained using a THO basis characterized by $b=0.7$ fm and $\gamma=1.4$ fm$^{-2}$, although we have checked that the resonance follows the same trend under other basis choices. This makes clear that quasibound states, such as the ground state in $^{26}$O, provide an ideal case for the stabilization approach. The calculated ground-state wave function, adjusted to the correct three-body energy, contains only $\sim 60$\% of $(d_{3/2})^2$ configurations, showing a large mixing with negative-parity components. As discussed in Refs.~\cite{Catara84,zhukov93}, this mixing favors the formation of a dominant dineutron arrangement in the wave function, as shown in Fig.~\ref{fig:26Onn} and already reported in the original calculations of Ref.~\cite{Hagino16}. 

The three-body decay energy spectrum for $^{26}$O, populated by knocking out a proton from $^{27}$F in inverse kinematics, can be obtained by starting from a reference wave function for the beam nucleus. This is achieved in Ref.~\cite{Hagino16} by using the Green's function method. As compared to $^{26}$O, the extra $d_{5/2}$ proton of $^{27}$F provides enough energy to bind the system. This is due to the tensor force between the valence proton and neutrons, which can be modeled in our three-body approach by modifying the effective spin-orbit strength in the $\text{core}+n$ potential (see Ref~\cite{Hagino14} and references therein). By fixing this strength to reproduce the energy of the $d_{3/2}$ bound state in $^{26}$F, the experimental two-neutron separation energy of $^{27}$F can be recovered. In the present calculations, the computed wave function for $^{27}$F has a dominant $(d_{3/2})^2$ character ($\sim 85$\%) and is concentrated at smaller distances due to the increased binding energy with respect to $^{26}$O.

\begin{figure}[t]
\centering
		\includegraphics[width=0.6\textwidth]{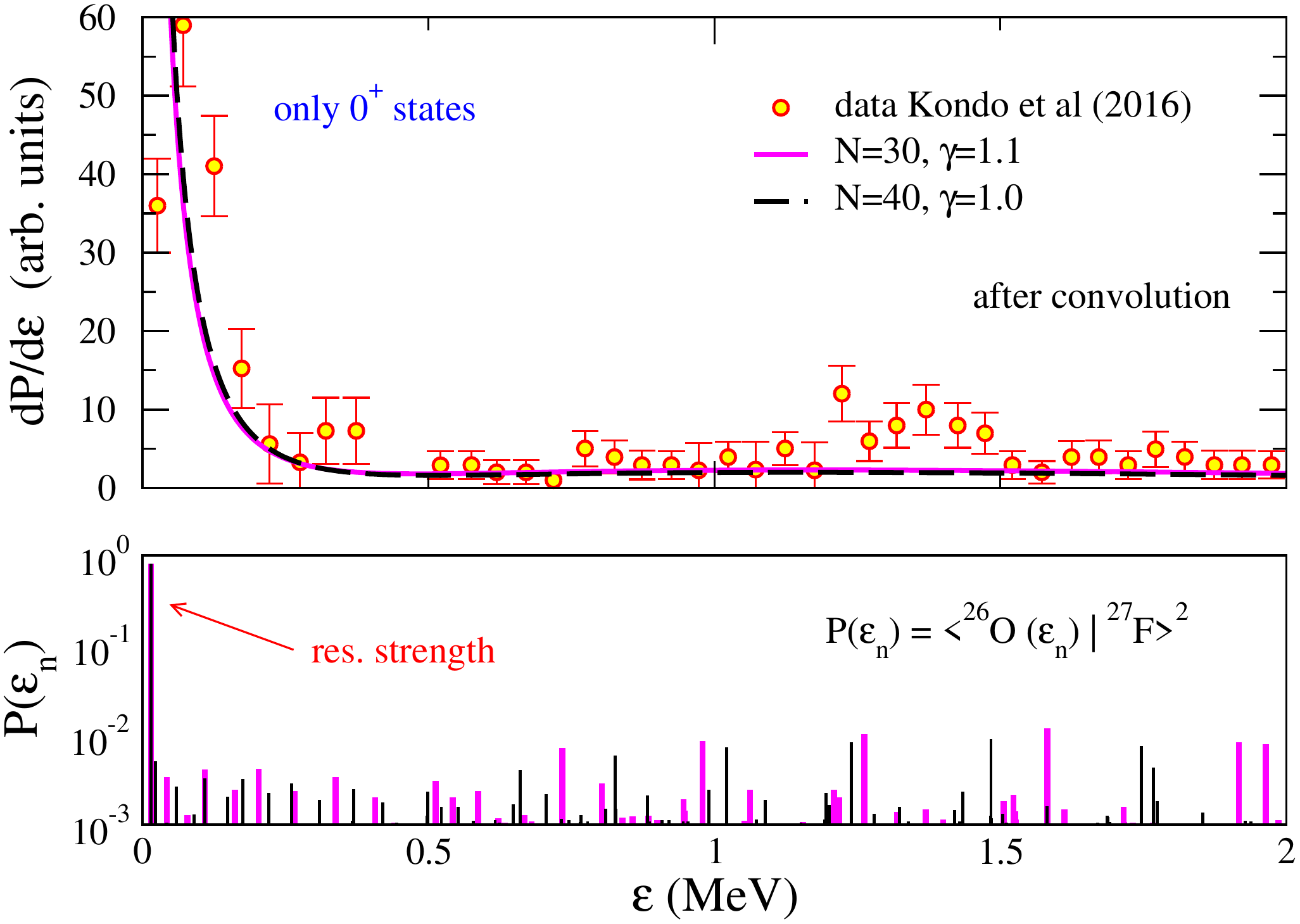}
		\caption{Three-body decay energy spectrum for $^{26}$O, compared with the available data from Ref.~\cite{Kondo16}. See text for details.} \label{fig:decay26O}
\end{figure}

Using these ingredients, in the present work we propose a simple estimation for the decay of $^{26}$O by assuming that the proton knockout from $^{27}$F is a sudden process. Under such assumption, the decay energy distribution is proportional to the square of the overlap function between the initial and final nuclei. In our discrete PS representation,
\begin{equation}
P(\varepsilon_n) = \langle {^{26}}\text{O}(\varepsilon_n)|{^{27}}\text{F} \rangle^2,
\end{equation}
such that the sum over all $\varepsilon_n$ values is normalized to unity. This yields a discrete spectrum which can be compared with experimental data after folding with the corresponding energy resolution. Note that, in order to achieve a detailed description of the low-energy continuum, it is now convenient to describe the states of $^{26}$O using a THO basis characterized by a smaller $\gamma$ parameter and a larger basis set $N$. Preliminary results, using two different bases, are shown in Fig.~\ref{fig:decay26O}. In the lower panel, it is shown that while the different discretizations provide rather different representations of the non-resonant continuum, the resonant strength (located just above the three-body threshold) is not affected. This is a consequence of the validity of the stabilization approach to describe low-lying resonances. Note that the computed resonant strength amounts for $\sim 80$\% of the total. In the upper panel, we compare the folded decay energy distribution with the data from Ref.~\cite{Kondo16}, normalized to the same total strength. The resonant peak associated to the 0$^+$ ground-state resonance is very well reproduced, and we show that the calculated spectra using different THO bases are equivalent. If we repeat the calculations without the $n$-$n$ interaction, the resonant peak is shifted to $\sim 1.5$ MeV, i.e., twice the $d_{3/2}$ single-particle level in the adopted model. This means that the neutron-neutron pairing is essential in defining the properties of $^{26}$O, as discussed in Refs.~\cite{Hagino14,Hagino16}. Calculations for the 2$^+$ state, as well as a full study of the $nn$ correlation and its evolution at the dripline crossing for $N=18$, are ongoing.

\section{Resonance identification}
The stabilization approach introduced in the preceeding section can be used to describe multichannel resonant states, understood as localized continuum structures, in a discrete, bound-state basis. It is, however, best suited for narrow resonances at low continuum energies. Moreover, stable eigenstates associated to resonances do not provide, a priori, information about the decay mechanism nor the decay width.
Recently, in Ref.~\cite{casal19} we proposed an alternative procedure to identify and characterize few-body resonances, based on the diagonalization of a resonance operator in a basis of Hamiltonian pseudostates,
\begin{equation}
\widehat{M}=\widehat{H}^{-1/2}\widehat{V}\widehat{H}^{-1/2}; ~~~~~~\widehat{M}|\psi\rangle = m |\psi\rangle; ~~~~~~ |\psi\rangle=\sum_n\mathcal{C}_n|n\rangle.
\label{eq:resop}
\end{equation}
As opposed to non-resonant continuum states, resonances are expected to be more localized in the range of the nuclear potential. Thus, resonant states can be identified from the eigenstates of $\widehat{M}$ corresponding to the lowest negative eigenvalues $m$. In this method, the expansion of the states in terms of energy eigenstates $|n\rangle$ allows us to build energy distributions and compute resonance widths from their time evolution. For details see Ref.~\cite{casal19}.

\begin{figure}[t]
	\centering
	\includegraphics[width=0.375\textwidth]{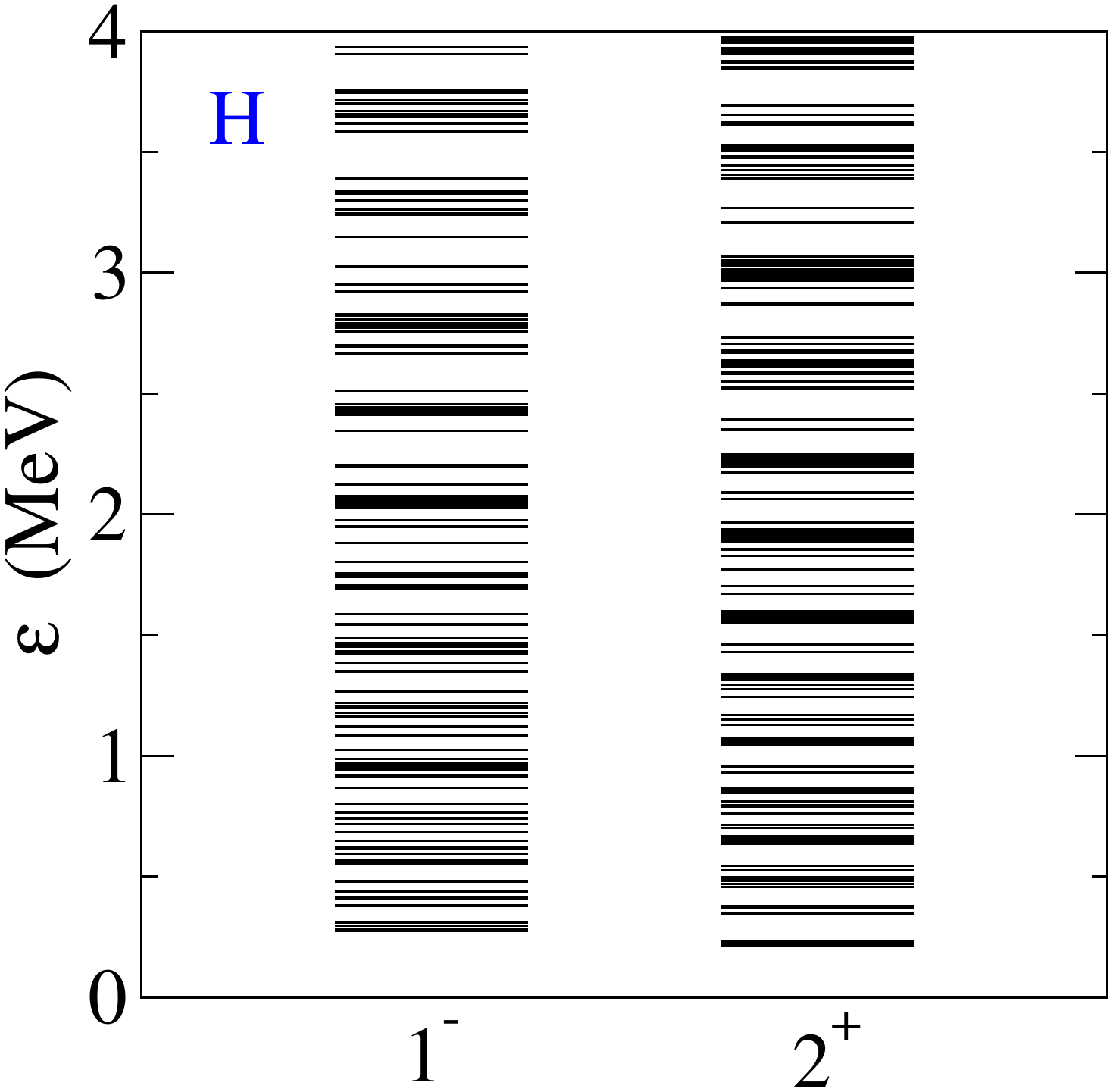}\hspace{15pt}\includegraphics[width=0.39\textwidth]{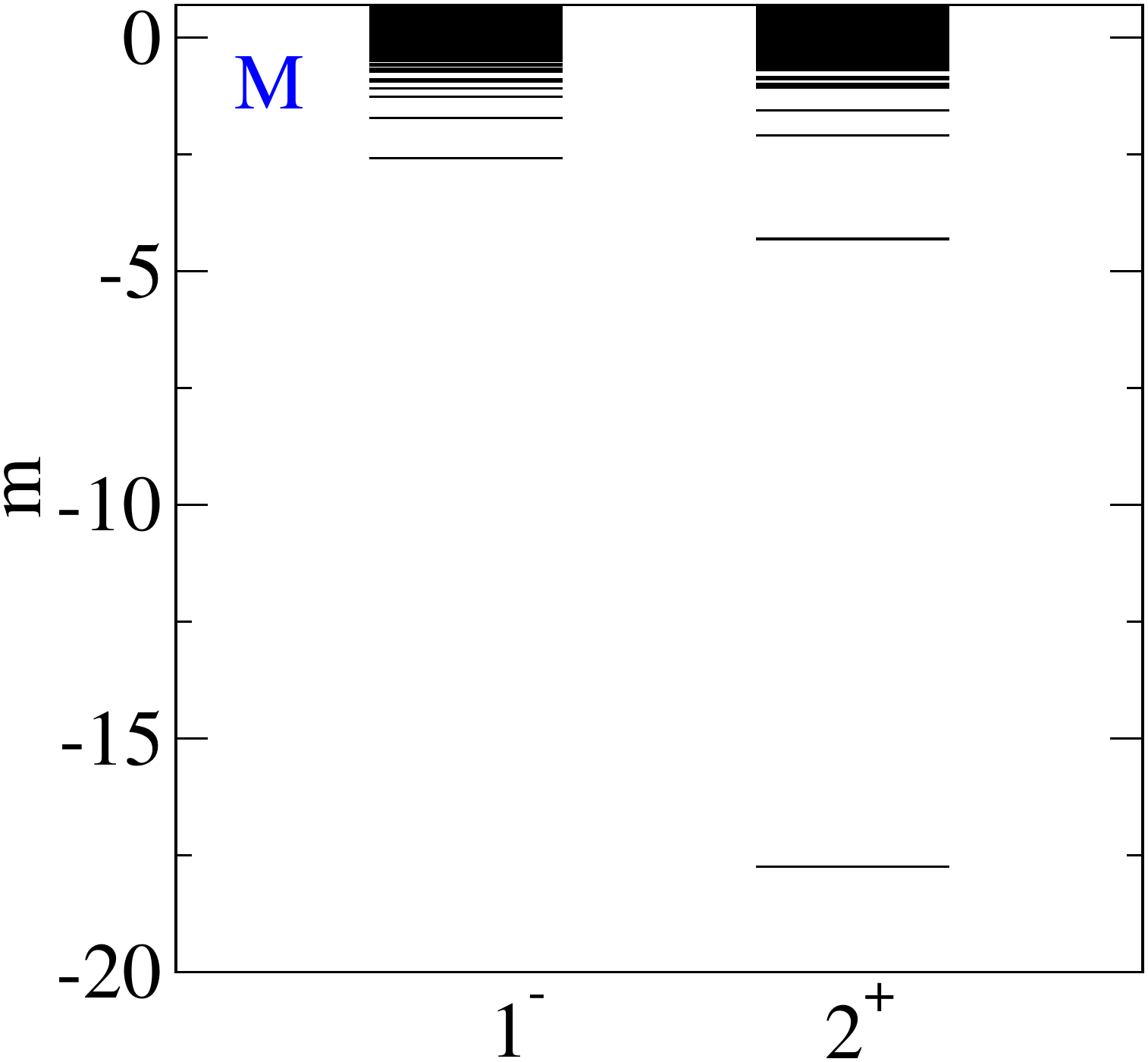}
	\caption{Example spectra of $\widehat{H}$ (left) and $\widehat{M}$ (right) for 1$^-$ and 2$^+$ states in $^6$He. Adopted from Ref.~\cite{casal19}.} \label{fig:H-M}
\end{figure}

As an example of the resonance identification, in Fig.~\ref{fig:H-M} we show the eigenstates of $\widehat{H}$ and $\widehat{M}$ for $j^\pi=1^-,2^+$ states in the two-neutron halo nucleus $^6$He described within a three-body $\alpha+n+n$ model~\cite{mroga05}. This system is known to exhibit a low-lying quadrupole resonance, which is clearly separated from the rest of continuum states and corresponds to a large negative value of $m$. Note that this state could not be trivially identified from the spectrum of $\widehat{H}$ in a large discrete basis.

\subsection{Two-neutron decay in $^{16}$Be}
The new method to identify and characterize few-body resonances was developed to address the two-neutron decay of the unbound system $^{16}$Be. This nuclear system was claimed to provide the first experimental observation of a correlated dineutron emission~\cite{spyrou12}. In Refs.~\cite{casal18,casal19}, we used the $^{14}\text{Be}+n$ potential employed in~\cite{lovell17} to construct $^{16}$Be wave functions in a three-body model. With this potential, the $^{15}$Be ground state is a $d_{5/2}$ resonance at 1.8 MeV above one-neutron emission. In our calculations, the 0$^+$ ground-state resonance of $^{16}$Be, adjusted to the known two-neutron separation energy of -1.35 MeV, was characterized by a width of 0.16 MeV. This value was found to be in good accord with the findings in Ref.~\cite{lovell17} that use the hyperspherical $R$-matrix method to compute the actual three-body scattering states, a fact which supports the realiability of the method to describe the properties of unbound three-body systems using a discrete basis. With the same Hamiltonian, we were able to predict also a 2$^+$ resonance at higher excitation energies, while no 1$^-$ resonance could be found by using the adopted model. An experimental confirmation of these predictions is still pending.

\begin{figure}[b!]
	\centering
	\includegraphics[width=0.48\textwidth]{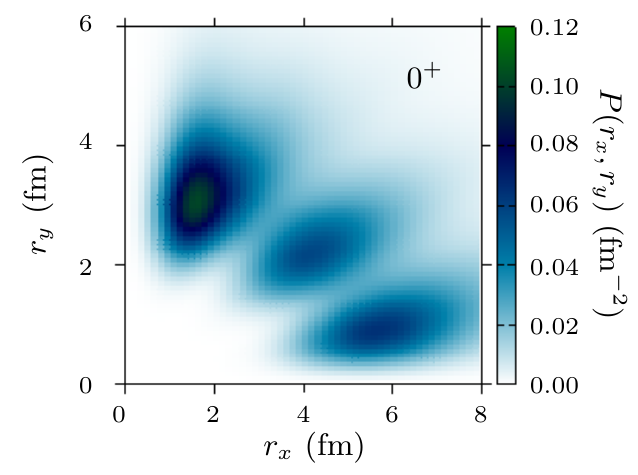}\hspace{10pt}\includegraphics[width=0.48\textwidth]{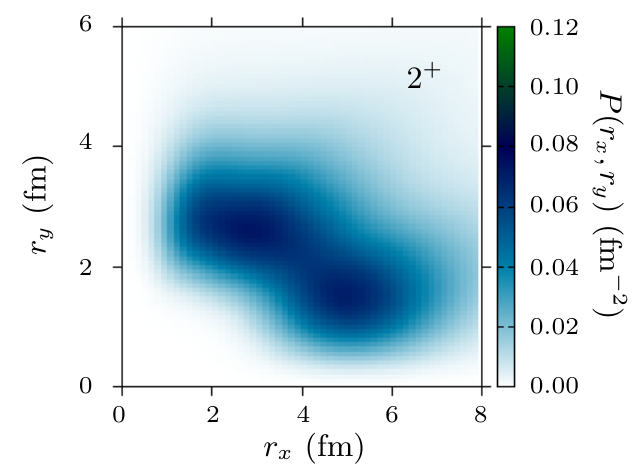}
	\caption{Probability density (fm$^{-2}$) for the 0$^+$ ground state (left) and 2$^+$ resonance (right) in $^{16}$Be, as a function of $r_x=r_{nn}$ and $r_y=r_{c-nn}$.} \label{fig:16Benn}
\end{figure}

Our three-body calculations for $^{16}$Be present a strong dineutron component for the 0$^+$ ground state, which is not dominant for the 2$^+$ state. This is shown in Fig.~\ref{fig:16Benn}, and would favor the picture of a correlated two-neutron emission from the ground state. The experimental signature for such a decay can be accessed from the $n$-$n$ relative energy distributions~\cite{spyrou12,marques19}. The theoretical interpretation of this observable could be achieved from the currents describing the flux of the resonance density escaping the potential well. Preliminary calculations on this line, compared with recent experimental data currently under analysis~\cite{marques19}, suggest that a strong signal at low relative energies might appear as a direct consequence of the $n$-$n$ interaction. A systematic study of the decay, including that of the 2$^+$ excited state, is ongoing and will be presented elsewhere.

\section{Conclusions}
We have presented recent results on the description of unbound states in three-body nuclear systems using a discrete-basis representation, the so-called pseudostate method. As an example, we have considered the low-lying dipole response in $^9$Be. Then, we have discussed the case of $^{26}$O and $^{16}$Be, which are two-neutron emitters. For $^{26}$O, we have shown that the stabilization approach is able to capture the properties of its barely unbound 0$^+$ ground state and the corresponding three-body decay-energy spectrum. In the case of $^{16}$Be, we have employed an identification method based on the eigenstates of a resonance operator, confirming the dominant dineutron component of its 0$^+$ ground state and predicting a new 2$^+$ resonance. The next steps will include a full study of the $n$-$n$ correlations in these systems and its role in shaping their decay properties.

\section*{Acknowledgements}

This work has been supported by the Dipartimento di Fisica e Astronomia, Università degli Studi di Padova (Italy) under project No.~CASA\_SID19\_1, by the Ministerio de Ciencia, Innovación y Universidades and FEDER funds (Spain) under projects No.~FIS2017-88410-P and~FPA2016-77689-C2-1-R, and by the European Union's Horizon 2020 research and innovation program under Grant No.~654002.



\bibliography{biblio.bib}

\begin{thebibliography}{10}
\providecommand{\url}[1]{\texttt{#1}}
\providecommand{\urlprefix}{URL }
\expandafter\ifx\csname urlstyle\endcsname\relax
  \providecommand{\doi}[1]{doi:\discretionary{}{}{}#1}\else
  \providecommand{\doi}{doi:\discretionary{}{}{}\begingroup
  \urlstyle{rm}\Url}\fi
\providecommand{\eprint}[2][]{\url{#2}}

\bibitem{tanihata13}
I.~Tanihata, H.~Savajols and R.~Kanungo,
\newblock \emph{Recent experimental progress in nuclear halo structure
  studies},
\newblock Progress in Particle and Nuclear Physics \textbf{68}(Supplement C),
  215  (2013),
\newblock \doi{10.1016/j.ppnp.2012.07.001}.

\bibitem{kikuchi16}
Y.~Kikuchi, K.~Ogata, Y.~Kubota, M.~Sasano and T.~Uesaka,
\newblock \emph{Determination of a dineutron correlation in borromean nuclei
  via a quasi-free knockout (p,pn) reaction},
\newblock Prog. Theor. Exp. Phys. \textbf{2016}(10), 103D03 (2016),
\newblock \doi{10.1093/ptep/ptw148}.

\bibitem{aksyutina13a}
Y.~Aksyutina, T.~Aumann, K.~Boretzky, M.~Borge, C.~Caesar, A.~Chatillon,
  L.~Chulkov, D.~Cortina-Gil, U.~D. Pramanik, H.~Emling, H.~Fynbo, H.~Geissel
  \emph{et~al.},
\newblock \emph{Momentum profile analysis in one-neutron knockout from
  borromean nuclei},
\newblock Physics Letters B \textbf{718}(4), 1309  (2013),
\newblock \doi{10.1016/j.physletb.2012.12.028}.

\bibitem{spyrou12}
A.~Spyrou, Z.~Kohley, T.~Baumann, D.~Bazin, B.~A. Brown, G.~Christian, P.~A.
  DeYoung, J.~E. Finck, N.~Frank, E.~Lunderberg, S.~Mosby, W.~A. Peters
  \emph{et~al.},
\newblock \emph{First observation of ground state dineutron decay:
  $^{16}\text{Be}$},
\newblock Phys. Rev. Lett. \textbf{108}, 102501 (2012),
\newblock \doi{10.1103/PhysRevLett.108.102501}.

\bibitem{kohley13}
Z.~Kohley, T.~Baumann, D.~Bazin, G.~Christian, P.~A. DeYoung, J.~E. Finck,
  N.~Frank, M.~Jones, E.~Lunderberg, B.~Luther, S.~Mosby, T.~Nagi
  \emph{et~al.},
\newblock \emph{Study of two-neutron radioactivity in the decay of
  $^{26}\text{O}$},
\newblock Phys. Rev. Lett. \textbf{110}, 152501 (2013),
\newblock \doi{10.1103/PhysRevLett.110.152501}.

\bibitem{oishi17}
T.~Oishi, M.~Kortelainen and A.~Pastore,
\newblock \emph{Dependence of two-proton radioactivity on nuclear pairing
  models},
\newblock Phys. Rev. C \textbf{96}, 044327 (2017),
\newblock \doi{10.1103/PhysRevC.96.044327}.

\bibitem{zhukov93}
M.~Zhukov, B.~Danilin, D.~Fedorov, J.~Bang, I.~Thompson and J.~Vaagen,
\newblock \emph{Bound state properties of \text{B}orromean halo nuclei:
  $^{6}\text{He}$ and $^{11}\text{Li}$},
\newblock Physics Reports \textbf{231}(4), 151  (1993),
\newblock \doi{10.1016/0370-1573(93)90141-Y}.

\bibitem{nielsen01}
E.~Nielsen, D.~Fedorov, A.~Jensen and E.~Garrido,
\newblock \emph{The three-body problem with short-range interactions},
\newblock Physics Reports \textbf{347}(5), 373  (2001),
\newblock \doi{10.1016/S0370-1573(00)00107-1}.

\bibitem{lovell17}
A.~E. Lovell, F.~M. Nunes and I.~J. Thompson,
\newblock \emph{Three-body model for the two-neutron emission of
  $^{16}\text{Be}$},
\newblock Phys. Rev. C \textbf{95}, 034605 (2017),
\newblock \doi{10.1103/PhysRevC.95.034605}.

\bibitem{nguyen12}
N.~B. Nguyen, F.~M. Nunes, I.~J. Thompson and E.~F. Brown,
\newblock \emph{Low-temperature triple-alpha rate in a full three-body nuclear
  model},
\newblock Phys. Rev. Lett. \textbf{109}, 141101 (2012),
\newblock \doi{10.1103/PhysRevLett.109.141101}.

\bibitem{tolstikhin97}
O.~I. Tolstikhin, V.~N. Ostrovsky and H.~Nakamura,
\newblock \emph{Siegert pseudo-states as a universal tool: Resonances,
  $\mathit{S}$ \text{Matrix, Green Function}},
\newblock Phys. Rev. Lett. \textbf{79}, 2026 (1997),
\newblock \doi{10.1103/PhysRevLett.79.2026}.

\bibitem{desc03}
P.~Descouvemont, C.~Daniel and D.~Baye,
\newblock \emph{Three-body systems with \text{Lagrange-mesh} techniques in
  hyperspherical coordinates},
\newblock Phys. Rev. C \textbf{67}, 044309 (2003),
\newblock \doi{10.1103/PhysRevC.67.044309}.

\bibitem{matsumoto04}
T.~Matsumoto, E.~Hiyama, M.~Yahiro, K.~Ogata, Y.~Iseri and M.~Kamimura,
\newblock \emph{Four-body \text{CDCC} analysis of \text{6H}e + \text{12C}
  scattering},
\newblock Nucl. Phys. A \textbf{738}, 471 (2004),
\newblock \doi{10.1016/j.nuclphysa.2004.04.089}.

\bibitem{mroga05}
M.~Rodr\'{\i}guez-Gallardo, J.~M. Arias, J.~G\'omez-Camacho, A.~M. Moro, I.~J.
  Thompson and J.~A. Tostevin,
\newblock \emph{Three-body continuum discretization in a basis of transformed
  harmonic oscillator states},
\newblock Phys. Rev. C \textbf{72}, 024007 (2005),
\newblock \doi{10.1103/PhysRevC.72.024007}.

\bibitem{casal14}
J.~Casal, M.~Rodr\'{\i}guez-Gallardo, J.~M. Arias and I.~J. Thompson,
\newblock \emph{Astrophysical reaction rate for $^{9}\text{Be}$ formation
  within a three-body approach},
\newblock Phys. Rev. C \textbf{90}, 044304 (2014),
\newblock \doi{10.1103/PhysRevC.90.044304}.

\bibitem{karataglidis05}
S.~Karataglidis, K.~Amos and B.~G. Giraud,
\newblock \emph{Local scale transformations and extended matter distributions
  in nuclei},
\newblock Phys. Rev. C \textbf{71}, 064601 (2005),
\newblock \doi{10.1103/PhysRevC.71.064601}.

\bibitem{casal18}
J.~Casal,
\newblock \emph{Two-nucleon emitters within a pseudostate method: The case of
  $^{6}\text{Be}$ and $^{16}\text{Be}$},
\newblock Phys. Rev. C \textbf{97}, 034613 (2018),
\newblock \doi{10.1103/PhysRevC.97.034613}.

\bibitem{HaziTaylor70}
A.~U. Hazi and H.~S. Taylor,
\newblock \emph{Stabilization method of calculating resonance energies: Model
  problem},
\newblock Phys. Rev. A \textbf{1}, 1109 (1970),
\newblock \doi{10.1103/PhysRevA.1.1109}.

\bibitem{TaylorHazi76}
H.~S. Taylor and A.~U. Hazi,
\newblock \emph{Comment on the stabilization method: Variational calculation of
  the resonance width},
\newblock Phys. Rev. A \textbf{14}, 2071 (1976),
\newblock \doi{10.1103/PhysRevA.14.2071}.

\bibitem{Descouvemont15}
P.~Descouvemont, T.~Druet, L.~F. Canto and M.~S. Hussein,
\newblock \emph{Low-energy $^{9}\text{Be}+^{208}\text{Pb}$ scattering, breakup,
  and fusion within a four-body model},
\newblock Phys. Rev. C \textbf{91}, 024606 (2015),
\newblock \doi{10.1103/PhysRevC.91.024606}.

\bibitem{Sumiyoshi02}
K.~Sumiyoshi, H.~Utsunomiya, S.~Goko and T.~Kajino,
\newblock \emph{Astrophysical reaction rate for $\alpha(\alpha
  n,\gamma){^9}\text{Be}$ by photodisintegration},
\newblock Nuclear Physics A \textbf{709}, 467  (2002),
\newblock \doi{10.1016/S0375-9474(02)01058-8}.

\bibitem{Arnold12}
C.~W. Arnold, T.~B. Clegg, C.~Iliadis, H.~J. Karwowski, G.~C. Rich, J.~R.
  Tompkins and C.~R. Howell,
\newblock \emph{Cross-section measurement of
  ${}^{9}\text{Be}$($\ensuremath{\gamma},n$)${}^{8}\text{Be}$ and implications
  for
  $\ensuremath{\alpha}+\ensuremath{\alpha}+n\ensuremath{\rightarrow}{}^{9}\text{Be}$
  in the $r$ process},
\newblock Phys. Rev. C \textbf{85}, 044605 (2012),
\newblock \doi{10.1103/PhysRevC.85.044605}.

\bibitem{RdDiego10}
R.~de~Diego, E.~Garrido, D.~V. Fedorov and A.~S. Jensen,
\newblock \emph{Relative production rates of \text{6He}, \text{9Be}, \text{12C}
  in astrophysical environments},
\newblock Eur. Phys. Lett. \textbf{90}, 52001 (2010),
\newblock \doi{10.1209/0295-5075/90/52001}.

\bibitem{Forseen03}
C.~Forss\'en, N.~B. Shul'gina and M.~V. Zhukov,
\newblock \emph{Radiative capture and electromagnetic dissociation involving
  loosely bound nuclei: The ${}^{8}\text{B}$ example},
\newblock Phys. Rev. C \textbf{67}, 045801 (2003),
\newblock \doi{10.1103/PhysRevC.67.045801}.

\bibitem{casal15}
J.~Casal, M.~Rodr\'{\i}guez-Gallardo and J.~M. Arias,
\newblock \emph{$^{9}\text{Be}$ elastic scattering on $^{208}\text{Pb}$ and
  $^{27}\text{Al}$ within a four-body reaction framework},
\newblock Phys. Rev. C \textbf{92}, 054611 (2015),
\newblock \doi{10.1103/PhysRevC.92.054611}.

\bibitem{Lunderberg12}
E.~Lunderberg, P.~A. DeYoung, Z.~Kohley, H.~Attanayake, T.~Baumann, D.~Bazin,
  G.~Christian, D.~Divaratne, S.~M. Grimes, A.~Haagsma, J.~E. Finck, N.~Frank
  \emph{et~al.},
\newblock \emph{Evidence for the ground-state resonance of $^{26}\mathbf{O}$},
\newblock Phys. Rev. Lett. \textbf{108}, 142503 (2012),
\newblock \doi{10.1103/PhysRevLett.108.142503}.

\bibitem{Kondo16}
Y.~Kondo, T.~Nakamura, R.~Tanaka, R.~Minakata, S.~Ogoshi, N.~A. Orr, N.~L.
  Achouri, T.~Aumann, H.~Baba, F.~Delaunay, P.~Doornenbal, N.~Fukuda
  \emph{et~al.},
\newblock \emph{Nucleus $^{26}\text{O}$: A barely unbound system beyond the
  drip line},
\newblock Phys. Rev. Lett. \textbf{116}, 102503 (2016),
\newblock \doi{10.1103/PhysRevLett.116.102503}.

\bibitem{Hagino16}
K.~Hagino and H.~Sagawa,
\newblock \emph{Decay dynamics of the unbound $^{25}\text{O}$ and
  $^{26}\text{O}$ nuclei},
\newblock Phys. Rev. C \textbf{93}, 034330 (2016),
\newblock \doi{10.1103/PhysRevC.93.034330}.

\bibitem{GPT}
D.~Gogny, P.~Pires and R.~D. Tourreil,
\newblock \emph{A smooth realistic local nucleon-nucleon force suitable for
  nuclear \text{Hartree-Fock} calculations},
\newblock Physics Letters B \textbf{32}(7), 591  (1970),
\newblock \doi{10.1016/0370-2693(70)90552-6}.

\bibitem{Catara84}
F.~Catara, A.~Insolia, E.~Maglione and A.~Vitturi,
\newblock \emph{Relation between pairing correlations and two-particle space
  correlations},
\newblock Phys. Rev. C \textbf{29}, 1091 (1984),
\newblock \doi{10.1103/PhysRevC.29.1091}.

\bibitem{Hagino14}
K.~Hagino and H.~Sagawa,
\newblock \emph{Correlated two-neutron emission in the decay of the unbound
  nucleus ${}^{26}\text{O}$},
\newblock Phys. Rev. C \textbf{89}, 014331 (2014),
\newblock \doi{10.1103/PhysRevC.89.014331}.

\bibitem{casal19}
J.~Casal and J.~G\'omez-Camacho,
\newblock \emph{Identifying structures in the continuum: Application to
  $^{16}\text{Be}$},
\newblock Phys. Rev. C \textbf{99}, 014604 (2019),
\newblock \doi{10.1103/PhysRevC.99.014604}.

\bibitem{marques19}
B.~Monteagudo and F.~M. Marqués,
\newblock private communication (2019).

\end{thebibliography}

\nolinenumbers

\end{document}